\documentclass{PoS}
\usepackage{natbib}

\title{Stellar Archaeology: New Science with Old Stars}

\ShortTitle{Stellar Archaeology: New Science with Old Stars}

\author{\speaker{Anna Frebel}\\
        Harvard-Smithsonian Center for Astrophysics, 60 Garden St, Cambridge, MA 02138, USA\\
        E-mail: \email{afrebel@cfa.harvard.edu}}

\abstract{The abundance patterns of metal-poor stars provide us a
        wealth of chemical information about various stages of cosmic
        chemical evolution. In particular, these stars allow us to
        study the formation and evolution of the elements, and the
        involved nucleosynthesis processes. This knowledge is
        invaluable for our understanding of the nature and condition
        of the early Universe, and the associated processes of early
        star- and galaxy formation. This proceeding summarizes the
        astrophysical topics and questions that can be addressed with
        metal-poor stars. For the full version of the review, the
        reader is referred to Frebel 2010.}

\FullConference{11th Symposium on Nuclei in the Cosmos, NIC XI\\
		July 19-23, 2010\\
		Heidelberg, Germany}

\begin{document}

\section{Introduction}
After the Big Bang, the first stars that formed from the pristine gas
  were very massive, of the order of
100\,M$_{\odot}$ (e.g., \citealt{bromm01, yoshida08}). After a very
short life time these co-called Population\,III stars exploded as
supernovae, which then provided the first metals to the still
primordial interstellar medium. All subsequent generations of stars,
Pop\,II, formed from chemically enriched material. The most metal-poor
stars are the earliest and most extreme Population\,II objects and
belong to the stellar generations that formed from the non-zero
metallicity gas left behind by the first stars.  

In their atmospheres these old objects preserve details of the
chemical composition of their birth gas cloud. They thus provide
stellar archaeological evidence of the earliest times of the
Universe. In particular, the chemical abundance patterns provide
information about the formation and evolution of the elements and the
involved nucleosynthesis processes and sites.  By extension,
metal-poor stars provide constraints on the nature of the first stars,
the initial mass function, and the chemical yields of first/early SNe.
This knowledge is invaluable for our understanding of the cosmic
chemical evolution and the onset of star- and galaxy formation
processes including the formation of the galactic halo. In summary,
galactic metal-poor stars are the local equivalent of the
high-redshift Universe, enabling observational constraints on the
nature of the first stars and supernovae, and more generally, on
various theoretical works on the early Universe.

Due to their low masses ($\sim0.8$\,M$_{\odot}$) metal-poor stars have
extremely long lifetimes that exceed the current age of the Universe
of $\sim14$\,Gyr  \citep{WMAP}. Hence, these stellar ``fossils''
of the early Universe are still observable.  However, the most
metal-poor stars (e.g., stars with\footnote{The main indicator used to
determine stellar metallicity is the iron abundance, [Fe/H], which is
defined as \mbox{[A/B]}$ = \log_{10}(N_{\rm A}/N_{\rm B})_\star -
\log_{10}(N_{\rm A}/N_{\rm B})_\odot$ for the number N of atoms of
elements A and B, and $\odot$ refers to the Sun. With few exceptions,
[Fe/H] traces the overall metallicity of the objects fairly well.}
$\mbox{[Fe/H]}<-5.0$; \citealt{HE1327_Nature}) are extremely rare
\citep{schoerck}, and hence difficult to identify. The most promising
way forward is to survey large volumes far out into the Galactic
halo. Past surveys include the HK survey and the Hamburg/ESO survey
\citep{ARAA} which have been very successful in producing large
samples of extremely metal-poor stars (with $\mbox{[Fe/H]}<-3.0$).  It
was also shown that there are many different types of abundance
patterns that arise as the result of specific nucleosynthesis
processes. The fact that the nuclear physics details of these
processes can be probed with the help of stars, means that stellar
astrophysics becomes ``nuclear astrophysics''. This is a very
complementary approach to experimental nuclear physics that is often
limited in its attempts to create the most exotic nuclei or extreme
processes, like the r-process, in the laboratory.

Over the past few years, a number of extensive reviews have been
published on the various roles and applications of metal-poor stars to
different astrophysical topics. \citet{ARAA} report on the discovery
history, search techniques and results, and the different abundance
``classes'' of metal-poor stars. \citet{sneden_araa} reviewed the
evolution of neutron-capture elements in the Galaxy, including the s-
and r-process stars. Recent advances regarding the stellar contents of
different types of dwarf galaxies are presented in
\citet{tolstoy_araa} and \citet{koch_biermann}. Finally,
\citet{frebel10} summarized the role of metal-poor stars in the
cosmological context, and how early star- and galaxy evolution can be
studied with them. This proceeding is based on \citet{frebel10}, and
thus only briefly summarizes a few specific aspects of what is
described in more detail in the above review.

\begin{figure*}  [!]
\begin{center}
\includegraphics[width=13cm,clip=true,bbllx=65,bblly=423,bburx=528,bbury=655]{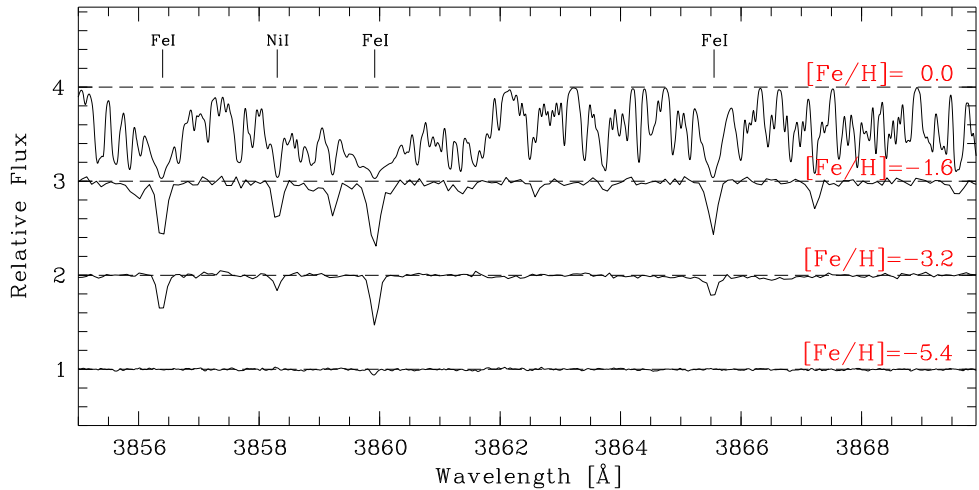}
  \caption{\label{spec_comp} Spectral comparison of stars in the
  main-sequence turn-off region with different metallicities. Several
  atomic absorption lines are marked. The variations in line strength
  reflect the different metallicities. From top to bottom: Sun with
  $\mbox{[Fe/H]}=0.0$, G66-30  with $\mbox{[Fe/H]}=-1.6$
  \citep{norris_emp3}, G64-12 with $\mbox{[Fe/H]}=-3.2$
  \citep{Aokihe1327}, and HE1327-2326 with $\mbox{[Fe/H]}=-5.4$
  \citep{HE1327_Nature}.  Reproduced from \citet{frebel10}.}
\end{center}
  \end{figure*}

\section{The Most Metal-Poor Stars}

The first star with a record-low iron abundance was found in 2001. The
faint ($V=15.2$) red giant HE~0107$-$5240 has $\mbox{[Fe/H]}=-5.2$
\citep{HE0107_Nature}. In 2004, the bright ($V=13.5$) subgiant
HE~1327$-$2326 was discovered \citep{HE1327_Nature,
Aokihe1327}. HE~1327$-$2327 has an even lower iron abundance of
$\mbox{[Fe/H]}=-5.4$. This value corresponds to $\sim1/250,000$ of the
solar iron abundance. Interestingly, the entire mass of iron in
HE~1327$-$2326 is actually 100 times less than that in the Earth's
core. A third star with $\mbox{[Fe/H]}=-4.75$ \citep{he0557} was found
in 2006. The metallicity of the giant HE~0557$-$4840 is in between the
two $\mbox{[Fe/H]}<-5.0$ stars and the next most metal-poor stars are
$\mbox{[Fe/H]}\sim-4.2$. Hence it sits right in the previously claimed
``metallicity gap'' (between $\mbox{[Fe/H]}\sim-4.0$ and
$\mbox{[Fe/H]}\sim-5.0$; e.g. \citealt{shigeyama}) showing that the
scarcity of stars below $\mbox{[Fe/H]}\sim-4.2$ has no physical cause
but is merely an observational incompleteness. All three objects were
found in the Hamburg/ESO survey \citep{frebel_bmps,hes4} making it the
so far most successful database for metal-poor stars.

The most striking features in both $\mbox{[Fe/H]}<-5.0$ stars are the
extremely large overabundances of CNO elements
($\mbox{[C,N,O/Fe]}\sim2$ to 4).  HE~0557$-$4840 partly shares this
signature by also having a fairly large [C/Fe] ratio.  Other elemental
ratios [X/Fe] are somewhat enhanced in HE~1327$-$2327 with respect to
the stars with $-4.0<\mbox{[Fe/H]}<-2.5$, but less so for the two
giants. No neutron-capture element is detected in HE~0107$-$5240 or
HE~0557$-$4840, whereas, unexpectedly, Sr is observed in
HE~1327$-$2326. Despite expectations, Li could not be detected in the
relatively unevolved subgiant HE~1327$-$2326. The upper limit is
$\log\epsilon ({\rm Li})<0.6$, where $\log\epsilon ({\rm A})$ =
$\log_{10}(N_{\rm A}/N_{\rm H})$ + 12. This is surprising, given that
the primordial Li abundance is often inferred from similarly unevolved
metal-poor stars \citep{ryan_postprim}. Furthermore, the upper limit
found from HE~1327$-$2326, however, strongly contradicts the WMAP
value ($\log\epsilon ({\rm Li})=2.6$) from the baryon-to-photon ratio
\citep{WMAP}. This may indicates that the star formed from extremely
Li-poor material.

\begin{figure*} 
\begin{center}
 \includegraphics[clip=true,width=10.cm,bbllx=40,bblly=448,bburx=456,bbury=740]{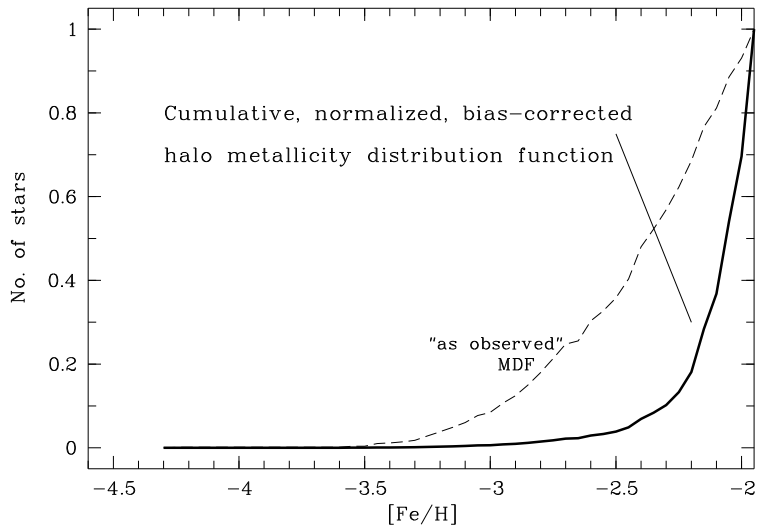}
 \caption{Metallicity distribution function of metal-poor halo
stars. Data taken from \citet{schoerck}. The dashed line shows the
``as observed'' function which reflects the distribution of stars
discovered in the Hamburg/ESO survey.  The solid line represents the
corrected function reflecting the actual underlying population of
metal-poor stars.}
\end{center}
\end{figure*}

HE~0107$-$5240 and HE~1327$-$2326 immediately became benchmark objects
to constrain various theoretical studies of the early Universe, such
as calculations of Pop\,III SN yields. Their highly individual
abundance patterns have been successfully reproduced by several
different SNe scenarios. This makes HE~0107$-$5240 and HE~1327$-$2326
early, extreme Pop\,II stars that possibly display the ``fingerprint''
of just one Pop\,III SN. \citet{UmedaNomotoNature} first matched the
yields of a faint 25\,M$_{\odot}$ SN that underwent a mixing and
fallback process to the observed abundances of HE~0107$-$5240.  To
achieve a simultaneous enrichment of a lot of C and only little Fe,
large parts of the Fe-rich SN ejecta have to fall back onto the newly
created black hole. Using yields from a SN with similar explosion
energy and mass cut, \citet{iwamoto_science} then reproduced the
abundance pattern of HE~1327$-$2326 also. Trying to fit the observed
stellar abundances, \citet{heger_woosley08} are employing an entire
grid of Pop\,III SN yields to search for the best match to the data. A
similar progenitor mass range as the \citet{UmedaNomotoNature}
25\,M$_{\odot}$ model was found to be the best match to have provided
the elemental abundances to the ISM from which these Pop\,II stars
formed.  \citet{meynet2005} explored the influence of stellar rotation
on elemental yields of 60\,M$_{\odot}$ near-zero-metallicity SNe.  The
stellar mass loss rate of rotating massive Pop\,III stars
qualitatively reproduces the CNO abundances observed in HE~1327$-$2326
and other carbon-rich metal-poor stars.

\section{Nuclear Astrophysics with Metal-Poor Stars}

About $5\%$ of metal-poor stars
with $\mbox{[Fe/H]}<-2.5$ contain a strong enhancement of
neutron-capture elements associated with the rapid (r-)
nucleosynthesis process \cite{ARAA} that is responsible for the
production of the heaviest elements in the Universe. In those stars we
can observe the majority (i.e., $\sim70$ of 94) of elements in the
periodic table: the light, $\alpha$, iron-peak, and light and heavy
neutron-capture elements. These elements were not produced in the
observed metal-poor star itself, but in a previous-generation
supernova explosions. We are thus able to study the ``chemical fingerprint'' of
individual supernova explosions that occurred just prior to the
formation of the observed star. So far, however, the nucleosynthesis
site of the r-process has not yet unambiguously been identified, but
supernovae with progenitor stars of $8-10$\,M$_{\odot}$ are the most
promising locations \cite{qian_wasserburg03}.

The giant HE~1523$-$0901 ($V = 11.1$) was found in a sample of bright
metal-poor stars \citep{frebel_bmps} from the Hamburg/ESO Survey. It
has the so far strongest enhancement in neutron-capture elements
associated with the r-process, $\mbox{[r/Fe]}=1.8$. Its metallicity is
$\mbox{[Fe/H]}=-3.0$ \citep{he1523}. The spectrum of HE~1523$-$0901
shows numerous strong lines of $\sim25$ neutron-capture elements, such
as those of Sr, Ba, Eu, Os, and Ir. A full discussion of the complete
abundance analysis will be given elsewhere (A.~Frebel et al. 2011, in
preparation). This makes possible a detailed study of the
nucleosynthesis products of the r-process. This fortuitously also
provides the opportunity of bringing together astrophysics and nuclear
physics because these objects act as a ``cosmic lab'' for both fields
of study.  The radioactive elements Th and U could be detected in this
star also (see the reviews of \citep{frebel_nic10, sneden_araa,
frebel10}). Only three stars have measured U abundances, of which
HE~1523$-$0901 has the most confidently determined value. From
comparing the stable Eu, Os, and Ir abundances with measurements of Th
and U, stellar ages can be derived.  Based on seven such chronometer
abundance ratios, the age of HE~1523$-$0901 was found to be
$\sim13$\,Gyr.

Next to Th and U, knowing all three abundances of Th, U, and Pb
provides a self-consistency test for r-process calculations. These
three abundances are intimately coupled, not only with each other but
also to the conditions (and potentially also the environment) of the
r-process. Hence, constraints on the different models yielding
different abundance distributions can be obtained by explaining the
stellar triumvirate of the Th, U and Pb abundances. Such constraints
lead to a better understanding of how and where r-process
nucleosynthesis can occur. In turn, improved r-abundance calculations
are crucial for reliably predicting the initial production ratios of
Th/r, U/r and Th/U, which are an implicit necessity for more accurate
age dating of r-process enhanced stars.  Based on new UVES data, a Pb
abundance could be determined for HE~1523$-$0901. The preliminary
abundance is $\log\epsilon(\rm Pb)\sim-0.35$. In summary, Pb is
produced through several decay channels. At the time when the
r-process event stops, there is the direct ($\beta$- and
$\beta$-delayed neutron) decay of very neutron-rich isobaric nuclei
with $A=206-208$ to $^{206}{\rm Pb}$, $^{207}{\rm Pb}$, and
$^{206}{\rm Pb}$. Then there is $\alpha$- and $\beta$-decay of nuclei
with $A\ge210$ back to Pb and finally the radioactive decay of the Th
and U isotopes back to Pb over the course of the age of the
Universe. The initial abundances of Th and U are driven in the same
way as Pb, i.e., by a direct channel of nuclei with $A=232$, 235 and
238, and an indirect way from the decay of r-process nuclei with
heavier masses. Of course, the Th and U abundances determine, in part,
the Pb abundance. Taking all these details into account when modeling
the r-process, the observed Pb abundances can be compared with model
predictions. After 13\,Gyr of decay, the prediction for the Th/U ratio
by \citet{frebel_ages} agrees well abundance measurement. With this
good level of agreement, HE 1523-0901 remains a vital probe for
observational nuclear astrophysics, which r-process models can
effectively be constrained.

\section{Extremely Metal-Poor Stars in Dwarf Galaxies}

\begin{figure*}[!t]
 \begin{center}
  \includegraphics[clip=true,width=15cm,bbllx=43, bblly=235, bburx=483,
   bbury=727]{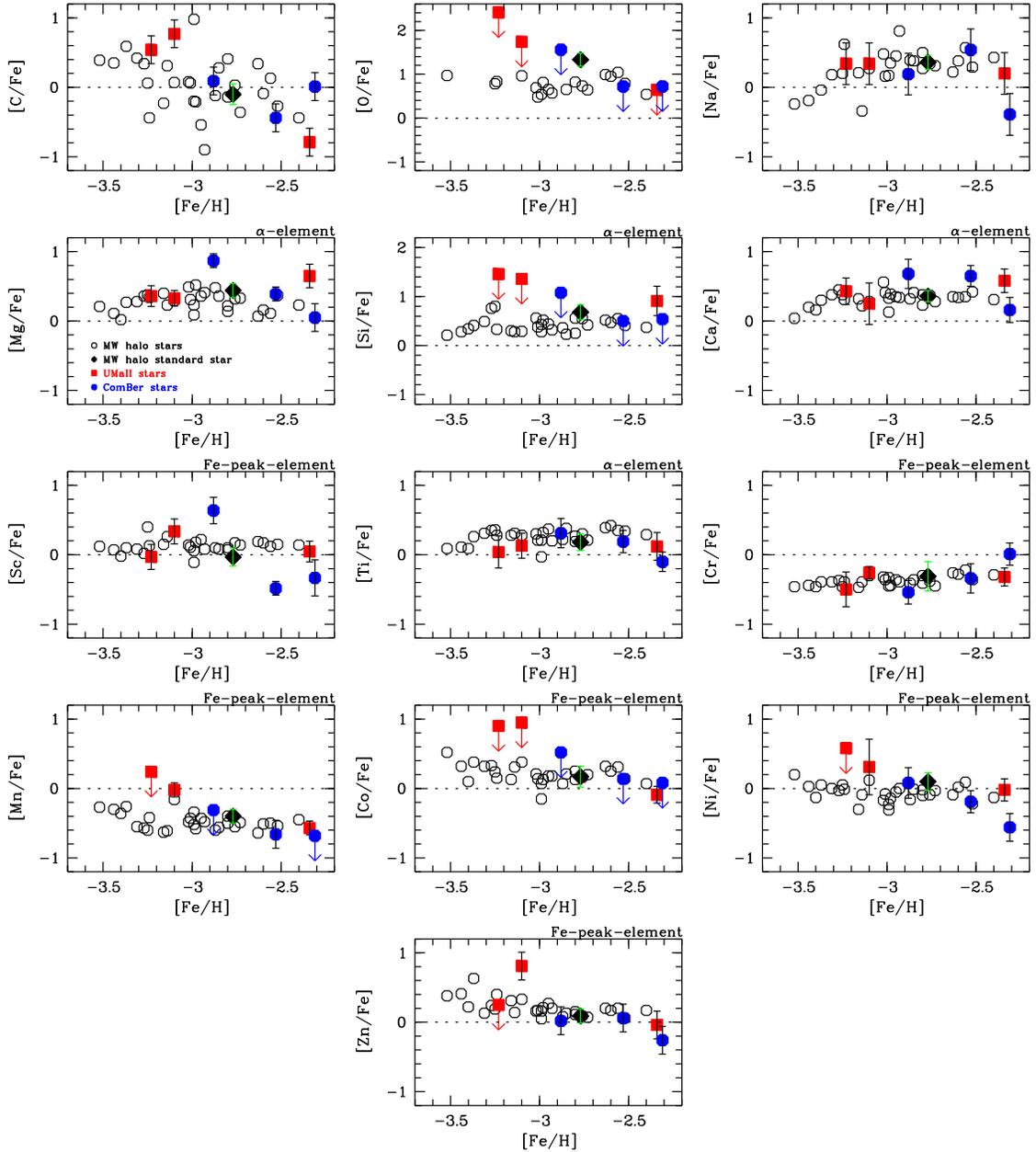} \caption{
     \label{cayrel_abundances_light} Abundances ([X/Fe]) of 
     ultra-faint dwarf galaxy stars for light and iron-peak elements
     in comparison with those of halo stars \citep{cayrel2004}
     (\textit{open black circles} ).  \textit{Red squares} indicate
     UMa\,II stars, \textit{blue circles} show ComBer
     stars. HD~122563, a halo ``standard'' star, is shown by
     a black diamond. From \citet{ufs}.}
 \end{center}
\end{figure*}

Simulations of the hierarchical assembly of galaxies within the cold
dark matter framework (\citealt{diemand07}; \citealt{springel}) show
that the Milky Way halo was successively built up from small dark
matter substructures, often referred to as galactic building
blocks. This hierarchical way of galaxy assembly had long been
suggested by \citet{sz78}. However, these simulations only include
dark matter, and it remains unclear to what extent small dark halos
contain luminous matter in the form of stars and gas. Studying the
onset of star formation and associated chemical evolution in dwarf
galaxies thus provides some of the currently missing information to
our understanding of how the observed properties of small satellites
relate to the dark matter substructures that build up larger galaxies.
The connection between the surviving dwarfs and those that dissolved
to form the halo is best addressed by examining in detail the stellar
chemical compositions of present-day dwarf galaxies. Establishing
detailed chemical histories of these different systems can provide
constraints on their dominant chemical enrichment mechanisms and
events, as well as the formation process of the Milky
Way. Specifically, detailed knowledge of the most metal-poor (hence,
oldest) stars in a given system allow insight into the earliest phases
of star formation before the effects of internal chemical evolution
were imprinted in stars born later with higher metallicity (see
reviews by \citealt{tolstoy_araa} and \citealt{koch_biermann}).

Assuming that the currently observable dwarf galaxies are analogs of
early systems that were accreted to form the halo, they provide an
opportunity to study this assembly history.  Particularly, the
metallicities of stars in dwarf galaxies must reach values as low as
(or lower) than what is currently found in the Galactic halo, and the
abundance ratio of those low-metallicity stars must be roughly equal
to those of equivalent stars in the halo. In the ``classical'' dSph
galaxies, higher-metallicity stars were found to have abundance ratios
different from comparable halo stars. Most strikingly, the
$\alpha$-element abundances are not enhanced, indicating different
enrichment mechanisms and timescales in these systems (e.g.,
\citealt{shetrone03}). But stars with $\mbox{[Fe/H]} < -3$ were
recently discovered the classical dwarf galaxy Sculptor dSph
\citep{kirby09}. A metallicity of $\mbox{[Fe/H]} = -3.8$ for one star
was confirmed from the high-resolution spectrum taken with
Magellan/MIKE \citep{scl}. Additional, similar objects have also been
found \citep{tafelmeyer10}. All their chemical abundances strongly
resemble those of halo stars with $\mbox{[Fe/H]} < -3.5$.

The metallicity-luminosity relationship of dwarf galaxies indicates
that the lowest luminosity dwarfs should contain extremely metal-poor
stars. \citet{kirby08} identified 15 extremely metal-poor stars in the
ultra-faint dwarfs, of which six were observed with high-resolution
spectroscopy (three located in each Ursa Major\,II and Coma Berenices
\citep{ufs}. Two of the stars in Ursa Major\,II are extremely
metal-poor having metallicities of $\mbox{[Fe/H]}<-3.0$. Additional
stars have been discovered in other ultra-faint dwarfs, in Segue and
Bootes \citep{geha09, norris10_booseg,norris_boo}, of which some have
been analyzed with high-resolution spectroscopy
\citep{norris10,norris10_seg,feltzing09}. The neutron-capture elements
are of extremely low abundances throughout the ultra-faint dwarf
galaxies, as well as in Sculptor at low metallicity. The Ba and Sr
values observed are well below the abundances found in typical MW halo
stars with similar Fe abundances. The low neutron-capture abundances
may represent a signature typical of stars with
$\mbox{[Fe/H]}\lesssim-2.0$ in dwarf galaxies. Similarly low values
have been found in the dSphs Hercules \citep{koch_her} and Draco
\citep{fulbright_rich}.

Overall, the detailed studies of stars in the ultra-faint dwarfs
provide the first evidence that the abundance patterns of light
elements ($Z < 30$) in the ultra-faint dwarfs, as well as in Sculptor
at $\mbox{[Fe/H]}\sim-3.8$, are remarkably similar to the Milky Way
halo.  The agreement renews support that the metal-poor end of the MW
halo population could have been built up from destroyed dwarf
galaxies.  These results above are broadly consistent with the
predictions of currently favored cosmological models
(\citealt{johnston08}).  The majority of the mass presumably in the
inner part of the stellar halo (at $\mbox{[Fe/H]}\sim-1.2$ to $-1.6$)
was formed in much larger systems such as the Magellanic Clouds.  New
results support a scenario where the ultra-faint dwarf galaxies
contributed some individual metal-poor stars that are now found
primarily in the outer Galactic halo (although not
exclusively). However, these systems may not have been sufficiently
numerous to account for the entire metal-poor end of the Fe
metallicity distribution of the Milky Way halo. Since the classical
dSphs contain more stellar mass and have been shown to also contain at
least some of the most metal-poor stars, they could have been a major
source of the lowest-metallicity halo content.

\section{Outlook}
Over the last few years many extremely metal-poor stars have been
discovered, particularly in the least-luminous dwarf galaxies. This
has opened a new window to study the formation of our Milky Way and
also small galaxies in the early Universe via detailed knowledge of
the chemical composition of individual stars. Consequently,
``near-field cosmology'' has received an enormous boost: Through the
availability of these new data as well as extensive theoretical
simulations on the same topic.

These early successes will be extended and underpinned by future
large-scale surveys such as the Australian Skymapper survey
\citep{keller} whose filter combinations have been designed
specifically for extracting information on the stellar parameters of
each star. This information is vital for characterizing the stellar
content of any galaxy, and invaluable for efficiently selecting large
numbers of metal-poor stars. No other survey will be able to match
these abilities.  Skymapper will provide an abundance of metal-poor
candidates in need for detailed high-resolution follow up to obtain
abundance measurements. By accessing large samples of fainter stars
in the outer Galactic halo and dwarf galaxies, the next big frontiers
in stellar archaeology and near-field cosmology can be tackled such as
the formation history of the halo, and the formation of the first
low-mass objects (e.g. \citealt{dtrans}). Estimates for the lowest
observable metallicity in the halo, $\mbox{[Fe/H]}=-7.3$ \citep{poll}
suggest that stars with such low metallicities can be found, most
likely in the outer halo. It thus appears within reach to find the
closest relatives to the first stars with the light-collecting power
of the next generation of optical telescopes, such as the Giant
Magellan Telescope, the thirty Meter Telescope or the European ELT, if
equipped with high-resolution spectrographs. Such facilities would
enable not only to reach out into the outer halo in search of the most
metal-poor stars, but also enable the aquisition of spectra with very
high-$S/N$ ratio of somewhat brighter stars. For example, for uranium
and lead measurements in r-process enhanced stars exquisite data
quality is required, which currently is only possible for the very
brightest stars.  \\\\ A.~F. acknowledges support through a Clay Fellowship
administered by the Smithsonian Astrophysical Observatory.


\end{document}